\documentclass[namedreferences]{SolarPhysics}
\usepackage[optionalrh]{spr-sola-addons} % For Solar Physics 
\usepackage{graphicx}                    % For eps figures, newer & more powerfull
\usepackage{color}                       % For color text: \color command
\usepackage{url}                         % For breaking URLs easily trough lines
\newcommand{\aap}{    {\it Astron. Astrophys.}}
\newcommand{\aaps}{   {\it Astron. Astrophys. Suppl.}}

\newcommand{\apjs}{   {\it Astrophys. J. Suppl.}}

\newcommand{\mnras}{  {\it Mon. Not. Roy. Astron. Soc.}}

\newcommand{\pasj}{   {\it Publ. Astron. Soc. Japan}}

\newcommand{\solphys}{{\it Solar Phys.}}

\begin{document}

%\begin{article}
\begin{opening}

\title{Solar Center--Limb Variation of the Strengths of Spectral Lines:
Classification and Interpretation of Observed Trends
}

%%%%%%%%%%%%%%%%%%%%%%%%%%%%%%%%%%%%%%%%%%%%%%%%%%%
%% Authors Names
%
\author{Y.~\surname{Takeda}$^{*,1,2}$ and 
        S.~\surname{UeNo}$^{3}$
       }

%%%%%%%%%%%%%%%%%%%%%%%%%%%%%%%%%%%%%%%%%%%%%%%%%%%
%% Runningheads
%
\runningauthor{Y. Takeda and S. UeNo}
\runningtitle{Center--Limb Variation of Solar Spectral Line Strengths}

%%%%%%%%%%%%%%%%%%%%%%%%%%%%%%%%%%%%%%%%%%%%%%%%%%%
%% Affiliations 
%
  \institute{$^{*}$ corresponding author, email: \url{takeda.yoichi@nao.ac.jp}\\
             $^{1}$ National Astronomical Observatory of Japan,
                    2-21-1 Osawa, Mitaka, Tokyo 181-8588, Japan\\
             $^{2}$ SOKENDAI, The Graduate University for Advanced Studies, 
                    2-21-1 Osawa, Mitaka, Tokyo 181-8588, Japan\\  
             $^{3}$ Kwasan and Hida Observatories, Kyoto University,
                     Kurabashira, Kamitakara, Takayama, Gifu 506-1314, Japan\\
             }

%%%%%%%%%%%%%%%%%%%%%%%%%%%%%%%%%%%%%%%%%%%%%%%%%%%
%%% Abstract 
\begin{abstract}
The equivalent widths ($W$) of 565 spectral lines in the wavelength range
of 4690--6870~\AA\ were evaluated at 31 consecutive points from the solar disk 
center ($\mu \equiv \cos\theta = 1$) to near the limb ($\mu = 0.25$) by applying 
the synthetic spectrum-fitting technique, in order to clarify the nature 
of their center--limb variations, especially the observed slope differing 
from line to line and its interpretation in terms of line properties.
We found that the distribution of the gradient $\beta$~$(\equiv -{\rm d}\log W/{\rm d}\log\mu)$ 
well correlates with that of ${\rm d}\log W/{\rm d}\log T$ index, which means that the 
center-to-limb variation of $W$ is determined mainly by the $T$-sensitivity 
of individual lines because the line-forming region shifts towards upper 
layers of lower $T$ as we go toward the limb. Further, the key to understanding 
the behavior of ${\rm d}\log W/{\rm d}\log T$ (depending on the temperature 
sensitivity of number population) is whether the considered species is in minor 
population stage or major population stage, by which the distribution of $\beta$ 
is explained in terms of differences in excitation potential and line strengths.
All the center--limb data of equivalent widths (as well as line-of-sight
turbulent velocity dispersions, elemental abundances, and mean line-formation 
depths derived as by-products) along with the solar spectra used for 
our analysis are made available as on-line materials.   
\end{abstract}

%%%%%%%%%%%%%%%%%%%%%%%%%%%%%%%%%%%%%%%%%%%%%%%%%%%
%% Keywords
%

\end{opening}
%-------------------------------------------------

%%%%%%%%%%%%%%%%%%%%%%%%%%%%%%%%%%%%%%%%%%%%%%%%%%%
%% Sections
%
% \section{}%\label{s:?} 

%Section 1. Introduction 
\section{Introduction}

Our Sun is the only star, the surface of which can be directly examined in detail. 
By making use of this advantage, we can get fruitful information on the 
physical structure of the solar atmosphere.

For example, since the widths of weak lines directly represent
the non-thermal velocity dispersion (macroturbulence) in the photosphere,
one can investigate its nature ({\it e.g.}, depth-dependence or angle-dependence) 
by studying how such the widths of various spectral lines vary from 
the disk center to the limb, as previously done mainly in the 1960--1970s 
(see, {\it e.g.}, Canfield, and Beckers, 1976, and the references therein).       

A similar situation holds true also for the line strengths (equivalent widths 
$W$ measured by integrating the normalized line-depth profile over wavelength), 
which contain a wealth of information because they are influenced by several 
factors ({\it e.g.}, abundance, microturbulence, and atmospheric structure) 
in a characteristic manner different from line to line. 
However, useful studies have been rather insufficient regarding the center-to-limb 
behavior in the equivalent widths of solar spectral lines:
\begin{itemize}
\item
Although several investigators derived the center--limb variations
of $W$ for representative spectral lines (for example, in the trial of studying 
the nature of microturbulence), the number of lines (typically, a few to 
$\approx$~10--20) as well as the observed points on the disk (mostly only 
several) are not necessarily sufficient ({\it e.g.}, Nissen, 1965; Gurtovenko, and 
Ratnikova, 1976; Kostik, 1982; Elste, 1986; Rodr\'{\i}guez Hidalgo, Collados, 
and V\'{a}zquez, 1994).
\item
Besides, available tables of equivalent widths of many solar spectral 
lines are mostly only for the disk center ({\it e.g.}, Moore, Minnaert, and 
Houtgast, 1966; Rutten, and van der Zalm, 1984a) or for the disk-integrated Sun 
(Rutten, and van der Zalm, 1984b; Meylan {\it et al.}, 1993), though 
Holweger (1967) derived $W$ values for hundreds of lines both at the disk 
center ($\mu \equiv \cos\theta=1$)\footnote{We define $\theta$ as the angle 
made by the line of sight and the normal to the surface at the observed point. 
Accordingly, $\theta = 0^{\circ}$ ($\cos\theta = 1$) at the disk center 
and $\theta = 90^{\circ}$ ($\cos\theta = 0$) at the limb. 
We often write $\cos\theta$ as $\mu$ in this paper.}
 and near the limb ($\mu=0.3$). Though Balthasar (1988) published $W$ 
values at 13 different points (from $\mu=1$ to $\mu = 0.112$) for 143 lines, 
they unfortunately appear to suffer systematic errors (see Section~3.3).
\item
From the viewpoint of public availability of the solar high-dispersion 
spectra, published extensive atlases are again limited to the disk center 
or the integrated Sun, except for that of Gurtovenko {\it et al.} (1975)
(profiles of 98 lines at 5 different $\mu$ are presented) and
Allende Prieto, Asplund, and Fabiani Bendicho (2004)
(spectra of 8 line regions at 6 $\mu$ points are provided as
electronic data).
\item
Few systematic studies on the trend/mechanism of center--limb variations 
of $W$ for different spectral lines ({\it e.g.}, how they depend on individual 
line properties) are available, though Jevremovi\'{c} {\it et al.} (1993) tried 
classification of the observed tendency by using 
Gurtovenko {\it et al.}'s (1975) line profile atlas.
\end{itemize}

In view of this circumstance, it may be worthwhile to revisit this problem 
based on new observational data and modern analysis techniques.
Recently, Takeda and UeNo (2017a; hereinafter referred to as TU17a) 
applied the semi-automatic spectrum-fitting method to the profiles of 86 lines 
observed at various points on the solar disk, in order to derive the line-of-sight
non-thermal velocity dispersion ($V_{\rm los}$), based on which the nature of 
photospheric macroturbulence was investigated. 
As a by-product of this profile-fitting analysis, the equivalent widths of 
the relevant lines could also be derived ({\it cf.} Section~4.3 in TU17a). 
Therefore, this revealed to be quite an efficient method for evaluation 
of $W$ values no matter how many observed points are involved.
Actually, in the subsequent paper of Takeda and UeNo (2017b; hereinafter TU17b)
which aimed at detecting the latitudinal dependence of surface temperature,
the center-to-limb variations of $W$ values for 28 spectral lines were
shown to be successfully established ({\it cf.} Section~3.1 therein).

Motivated by this experience, we decided to conduct a comprehensive study
on the solar center--limb variations of equivalent widths for a large number
of spectral lines based on the observational data of our own (new spectra 
obtained in our recent 2017 observations, in addition to the 2015 data 
used in TU17a). The objectives of this investigation are as follows.
\begin{itemize}
\item
How do the $W$ values of various spectral lines of different species 
behave themselves over the solar disk ({\it e.g.}, whether increasing or 
decreasing toward the limb)?
\item 
Is there any relation between the observed trend and line properties 
(such as ionization stage, excitation potential, line strength)? If so, 
is it possible to find a reasonable physical interpretation?
\end{itemize}

Also, such solar $W$ {\it vs.} $\mu$ data compiled for a large number of spectral 
lines may serve as useful database for empirically checking the validity 
of line-formation calculations, such as determining appropriate collision 
cross sections in non-LTE calculations ({\it e.g.,} Allende Prieto, {\it et al.}, 2004; 
Pereira, Asplund, and Kiselman, 2009; Takeda, and UeNo, 2015) or verifying 
state-of-the-art 3D hydrodynamical line formation models ({\it e.g.,} Lind, et al., 2017), 
which would be beneficial for both solar and stellar physics.

The remainder of this article is organized as follows. 
Our observations and data reduction are described in Section~2. 
We explain the procedures of our spectrum-fitting analysis 
for deriving the equivalent widths in Section~3.  
Section~4 is devoted to discussing the resulting center-to-limb
variations of $W$ values, where we will show that the observed trend
is reasonably interpreted in terms of the $T$-sensitivity of $W$.
We also compare our results with those of the past literature.
The conclusions are summarized in Section~5.
In addition, an appendix is presented, where our supplementary 
materials available on-line are described.

%Section 2. Observational Data (Fig. 1, Table !, Table 2)
\section{Observational Data}

Our observations were carried out in three periods (2015 November 3--5, 
2017 July 17--21, and 2017 October 31--November 3; JST) by using 
the 60~cm Domeless Solar Telescope (DST) with the Horizontal Spectrograph 
at Hida Observatory of Kyoto University (Nakai, and Hattori, 1985).
The configuration and surface appearance of the Sun in each period
are shown in Figure~1, which shows that an appreciable active region
existed slightly northward of disk center (which passed the meridian
on 2015 November 4) in 2015 November observation while the solar surface 
was quite clean in both 2017 July and 2017 November observations.
As to the positions targetted on the solar disk, we selected 32 
points on the northern meridian line of the solar disk 
(from the disk center to 0.97~$R_{0}$ with a step of 
30$''\approx$~0.03~$R_{0}$, where $R_{0}$ is the apparent radius 
of the solar disk), at which the slit was aligned along the E--W direction.
In this arrangement, the disk center and the nearest-limb point 
correspond to $\cos\theta =1$ and $\cos\theta = 0.25$, respectively, 
which means that the range of $\theta$ covered by our data is 
$0^{\circ}\le \theta \lesssim 76^{\circ}$ ({\it cf.} Table~1).

In the adopted setting of the spectrograph, a solar spectrum covering 
$\approx$~150$''$ (spatial) and $\approx$~21--24~\AA\ (wavelength) was recorded on 
the CCD detector with $1\times2$ binning (1600 pixels in the dispersion direction 
and 600 pixels in the spatial direction) by one-shot observation. 
After the whole set (consecutive observations on 32 points along the center-to-limb 
meridian) had been completed, we changed the central wavelength one after another.
By combining all the spectra, we could eventually accomplish the whole spectral 
range covering 4690--6870~\AA\ ({\it cf.} Table~2).
 
After reduction of the raw data (dark subtraction, spectrum extraction) 
carried out by following the standard procedure, the 1-dimensional spectrum was 
extracted by integrating over 200 pixels ($\approx 50''$; {\it i.e.}, $\pm 100$ 
pixels centered on the target point) along the spatial direction, 
which means that our spectrum corresponds to the spatial mean of 
a $\approx$~ 50$''$-size region including several tens of granular cells 
(typical size is on the order of $\approx$~1$''$).
Then, the continuum normalization was done by using the ``continuum'' 
task of IRAF\footnote{
  IRAF is distributed by the National Optical Astronomy Observatories,
  which is operated by the Association of Universities for Research
  in Astronomy, Inc. under cooperative agreement with
  the National Science Foundation.} (Image Reduction and Analysis Facility).
Regarding the wavelength calibration, we derived the wavelength {\it vs.} pixel 
relation (approximated by a second-order polynomial) for each region
by comparing our disk-center spectrum with that of the solar FTS 
(Fourier Transform Spectrometer) spectrum atlas (Neckel, 1994; Neckel, 1999), 
and the same relation was applied for the spectra of all points 
on the disk. Note that our wavelength calibration is not so 
accurate because of such an approximate treatment, and thus uncertainties 
of $\approx$~0.01--0.02~\AA\ are possible (see also Appendix~A.2).
Finally, the effect of scattered light was corrected according to 
the procedure described in Section~2.3 of Takeda and UeNo (2014).
The value of $\alpha$ (fraction of scattered light) we estimated
and adopted was 0.10 for most cases, except for the $\lambda > 5500$~\AA\ spectra 
obtained in the 2015 November observation, for which we found a slightly larger 
value of 0.15 to be appropriate.
The S/N ratios of the resulting spectra are sufficiently high (several 
hundreds or more), and the spectrum resolving power is $R \approx 140\,000$ 
({\it cf.} Section~4.2 in TU17a)
All the final center--limb spectra at each wavelength region, which we used
for our analysis, are provided as supplementary materials available on-line 
({\it cf.} Appendix~A).

%Section 3. Spectrum Fitting and Evaluation of Equivalent Widths (Figure 2, Figure 3, Table 3)
\section{Spectrum Fitting and Evaluation of Equivalent Widths}

\subsection{Line-Profile Modeling and Parameter Determination}

The modeling of line-profiles and the spectrum-fitting analysis 
was done in almost the same manner as described in TU17a.
The intensity profile $I(v,\theta)$\footnote{In Equations 1 and 3, the profile 
point is specified by $v$ (velocity variable) for simplicity, instead of $\lambda$ (wavelength).}
 emergent to direction angle $\theta$ is expressed as
\begin{equation}
I(v,\theta) = I^{0}(v,\theta) \otimes K(v) \otimes P(v),
\end{equation}
where `$\otimes$'' means the convolution.
Here, $I^{0}(v,\theta)$ is the intrinsic profile of outgoing specific intensity 
at the surface, which is written by the formal solution of radiative transfer as 
\begin{equation}
I^{0}(\lambda; \theta) = \int_{0}^{\infty} S_{\lambda}(t_{\lambda}) 
\exp(-t_{\lambda}/\cos\theta){\rm d}(t_{\lambda}/\cos\theta),
\end{equation}
where $S_{\lambda}$ is the source function and $t_{\lambda}$ is the optical depth
in the vertical direction. 
Further, $K(v)$ is the Gaussian macroturbulence broadening function
\begin{equation}
K(v) \propto \exp[-(v/V_{\rm los})^2],
\end{equation} 
$P(v)$ is the Gaussian instrumental profile with 
FWHM of 2.14~km~s$^{-1}$ (corresponding to $R\approx 140\,000$).

Regarding the calculation of $I^{0}(\lambda; \theta)$, we adopted Kurucz's 
(1993) ATLAS9 solar photospheric model ($T_{\rm eff}$ = 5780~K and $\log g = 4.44$) 
with a microturbulent velocity of $\xi$ = 1~km~s$^{-1}$ while assuming LTE.
We adopted the algorithm described in Takeda (1995) to search for 
the best-fit theoretical profile at each point ($\theta$) of the solar
disk, while varying three parameters 
[$\log\epsilon$ (elemental abundance), $V_{\rm los}$ (line-of-sight 
velocity dispersion), and  $\Delta\lambda_{\rm r}$ (wavelength shift)] 
for this purpose. As to the atomic parameters of each spectral line 
($gf$ values, damping constants), we exclusively adopted the values 
presented in Kurucz and Bell's (1995) compilation. 
The background opacities were included as fixed by assuming 
the solar abundances; {\it i.e.}, the opacities of other nearby lines 
were computed again with the atomic line data of Kurucz and Bell 
(1995) and the local continuum opacity were calculated according 
to Kurucz's (1993) ATLAS9 program.

After the solution has been converged, we can use the resulting abundance
solution ($\log\epsilon (\theta)$) to compute the corresponding equivalent 
width ($W(\theta)$) and the mean depth of line formation 
($\langle \log \tau (\theta )\rangle$) with the help of Kurucz's (1993) WIDTH9 program:
\begin{equation}
W(\theta) \equiv \int R^{0}_{\lambda}(\theta) {\rm d}\lambda
\end{equation}
and
\begin{equation}
\langle \log \tau (\theta )\rangle \equiv \frac{\int R^{0}_{\lambda}(\theta)\log 
 \tau_{5000}(\tau_{\lambda} = \cos\theta) {\rm d}\lambda} {\int R^{0}_{\lambda} 
(\theta) {\rm d}\lambda}
\end{equation}
where $\tau_{5000}$ is the continuum optical depth at 5000~$\rm\AA$,
$R^{0}_{\lambda}(\theta)$ is the line depth of theoretical 
intrinsic profile with respect to the continuum level 
[$R^{0}_{\lambda}(\theta) \equiv 1 - I^{0}_{\lambda}(\theta)/I^{0}_{\rm cont}(\theta)$]
and integration is done over the line profile.

% Table 3
\subsection{Selected Spectral Lines and Results}

In preparing the candidate list of possibly usable spectral lines, 
we mainly consulted the line list of Meylan {\it et al.} (1993), 
who extensively published the solar flux equivalent widths of 
around 570 carefully chosen lines. In addition, we also drew upon 
the work of Nissen (1965), Gurtovenko and Ratnikova (1976), 
Kostik (1982), and Balthasar (1988). Checking the profile of 
each line (in the disk-center spectrum) by eye while examining the 
theoretical strengths of the lines (existing in the neighborhood)
computed with the help of Kurucz and Bell's (1995) atomic line data, 
we required the condition that the feature is practically dominated 
only by the main line ({\it i.e.}, unaffected by significant blending 
of other lines) at least in the line core region.
In this eye-inspection process, the lower and upper wavelength 
limits of the profile-fitting range [$\lambda_{1}$, $\lambda_{2}$] were 
also determined. As a result, we could sort out $\approx$~600 high-quality 
spectral lines, to which our spectrum fitting analysis was applied.

It then turned out that the convergence was successfully attained for 565 lines, 
though the solution sometimes failed to converge (for $\approx$5\% of them).
The fundamental atomic data of these 565 spectral lines, the disk-center 
abundance solutions ($\log\epsilon_{00}$), and the equivalent widths at the disk 
center ($W_{00}$) and the limb ($W_{31}$) are presented in Table~3, 
while ``tableE.dat'' (which includes more detailed information of atomic data 
for each spectral line) is also presented as a supplementary data table.
Besides, the full results of $\log\epsilon(\theta)$, $W(\theta)$, $V_{\rm los}(\theta)$,
and $\langle \log \tau(\theta) \rangle$ (at each $\theta$ point) for all 565 lines
are also given as online material (see Appendix~A). 

Some remark may be appropriate here regarding the meaning of such 
derived equivalent widths ($W$) in view of the line blending effect. 
While $\log\epsilon (\theta)$ was derived from synthetic 
spectrum fitting by including background opacities of other lines in 
the neighborhood, the $W(\theta)$ inversely computed by using 
$\log\epsilon (\theta)$ corresponds to the pure contribution of 
the relevant (single) line. Accordingly, in case that any 
significant opacity-overlapping of other lines exists in the
theoretical spectrum synthesis, such derived $W(\theta)$ would be 
more or less different from the empirically evaluated equivalent 
width obtained by direct integration of the blended feature. 
However, since seriously blended lines are not included in our 
target lines because of our pre-check in advance (as explained 
above), our $W(\theta)$ values are regarded as almost equivalent 
to the empirically derived ones.
In a different context, there are some cases where appreciable contamination 
of other line actually exists (especially for weak-line cases) but 
its blending opacity was not included in the fitting of synthesized 
spectrum. For example, two forbidden lines [O~{\sc i}]~5577 and 
[O~{\sc i}]~6363 ({\it cf.} Table~3) are actually overlapped with 
weak molecular lines (C$_{2}$ lines for the former and CN line for 
the latter; {\it cf.} Mel\'{e}ndez, and Asplund, 2008; Takeda, 
{\it et al.}, 2015) Since molecular lines were not considered in this 
study, spectrum fitting as well as the following derivation of equivalent 
widths were done without taking into account these blended lines. 
Yet, even in such cases, the finally resulting $W$ values surely 
represent the actual equivalent widths of the total [O~{\sc i}]+C$_{2}$ 
or [O~{\sc i}]+CN feature (because satisfactory fitting could be 
accomplished by considering oxygen lines only). But they should not 
be regarded as corresponding to the pure contribution of [O~{\sc i}].

% Figure 2, Figure 3
\subsection{Consistency Check of Equivalent Widths}

As a check, we also measured the equivalent widths of all lines based on the spectra 
at point 00 (disk center: $\mu=1.0$) and point 31 (nearest to limb: $\mu = 0.25$) 
by using the conventional Gaussian-fitting method, in order to see whether 
they match those derived by Equation~4. It was then found that the agreement 
is fairly good as long as $W \lesssim 100$~m\AA. However, a systematic deviation 
begins to appear when $W$ exceeds $\approx100$~m\AA\ (in the sense that 
$W$(Gaussian) tends to be underestimated) and this tendency is more manifest
at the limb (Figure~2a$^{\prime}$) than at the disk center (Figure~2a).
This is because the damping wing becomes appreciable at $W \gtrsim 100$~m\AA;
and this effect is more apparent at the limb because lines get generally 
stronger there due to the lowered temperature in the line-forming region
(see Figures~2c and ~2c$^{\prime}$ for the representative case of 
Fe~{\sc i} 5434.523 line).

Comparisons of our $W$ values with those derived by four representative studies 
are shown in Figure~3, from which we can read the following tendency:
\begin{itemize}
\item
We can see a reasonable consistency between our equivalent widths
with those of Moore, Minnaert, and Houtgast (1966) (disk center; {\it cf.} Figure~3a),
of Rutten and van der Zalm (1984a) (disk center; {\it cf.} Figure~3b),
and of Holweger (1967) (disk-center and limb; {\it cf.} Figures~3c and 3d).
\item
However, Balthasar's (1988) equivalent widths apparently disagree ({\it i.e.}, 
larger by up to several tens of \%) with the values measured in this study for both 
at the disk center (Figure~3e) and the limb (Figure~3f), which makes us suspect 
the existence of appreciable systematic errors in his measurement.   
This was somewhat unexpected, because the FTS spectra (on which his 
study was based) are generally considered to be superior 
because they are unaffected by any scattered light. Since nothing is described 
in his paper regarding how the equivalent widths were measured, we have no 
idea about the reason for this systematic discrepancy. As seen from the tendency 
that the relative differences ($\Delta W/W$) tend to increase as lines 
become weaker ({\it cf.} the lower panels of these two figures), 
his continuum position might have been placed too high.
\end{itemize}

%Section 4. Discussion
\section{Discussion} 

\subsection{Classification of Spectral Lines}

Before we discuss the center-to-limb variation trend of the equivalent
widths of various lines derived in Section~3, it is worthwhile to divide 
them into two categories depending on the population status 
related to the condition of ionization equilibrium, as done in TU17b
(see Section~4.1 therein).
That is, if the relevant species of a spectral line belongs to the stage 
where its number population is small or negligible in the typical line-forming 
region ({\it e.g.}, $\tau \approx 10^{-1}$) compared to the total number of 
atoms for the element, we call it ``minor population species.''
On the other hand, if the considered species is in the dominant stage of the 
ionization equilibrium, we refer to this line as of ``major population species.''  
This classification can also be roughly done from the ionization potential 
($\chi^{\rm ion}$): As a rule of thumb for the solar case, a neutral species is of 
minor population if $\chi^{\rm ion} \lesssim 8$~eV, while it is of major population 
if $\chi^{\rm ion} \gtrsim 10$~eV.
Concerning the spectral lines we analyzed, almost all neutral species of heavier 
elements ($Z > 10$: $Z$ is atomic number) are of minor population, though light 
elements of neutral carbon (C~{\sc i}) and oxygen (O~{\sc i}) are of major population. 
Meanwhile, all ionized species are of major population. Only one indefinite case 
is Zn~{\sc i} ($\chi^{\rm ion} = 9.39$~eV), which is neither minor 
nor major, because the neutral and once-ionized populations of zinc are almost 
of the same order in the solar photosphere. The classification of each species 
is given in Table~3.

% Figure 4, Figure 5
\subsection{Center--Limb Variations of Line Strengths}

We are now ready to examine how the line strengths vary as we go from the 
disk center to the limb. Do they increase or decrease? 
The ratios of $\log (W/W_{00})$ (logarithm of the equivalent width
normalized by the disk-center value) are plotted against $\cos\theta$
in Figure~4a--4d (lines of minor population stage) and Figure~4e--4h 
(lines of major population stage), where each panel corresponds 
to different line-strength ($W_{00}$) group and different colors are used 
depending on whether the relevant key potential energy  
($\chi^{\rm ion} - \chi_{\rm low}$ for the minor population species and 
$\chi_{\rm low}$ for the major population species; see Section~4.3) 
is larger or smaller than 5~eV. We can grasp from these figures 
the rough outline of center--limb variations of $W$ and their 
line-by-line differences:
\begin{itemize}
\item
Most lines tend to be strengthened toward the limb ({\it i.e.}, 
$\log (W/W_{00}) > 0$). More precisely, all lines of minor population 
species as well as a large fraction of lines of major population species 
follow this tendency.
\item 
However, a small fraction of lines do show a progressive decrease 
with a decrease in $\cos \theta$, all of which are major population
lines of large $\chi_{\rm low}$ without exception. 
\item
These specific trends are apparently more manifest for the cases 
of larger potential energy (pink lines in Figure~4).
\item
It appears that the center--limb variations tend to become flatter 
as lines get stronger.
\end{itemize}

Let us further investigate these features more in detail toward a more 
definite classification and better understanding of the underlying mechanism.
To be more specific, we display the $\log W$ {\it vs.} $\cos\theta$ relations
of various representative lines in Figure~5, which are appropriately arranged 
for each species or assorted into groups suitable to study the effect of 
line parameters ({\it e.g.}, excitation potential or line strength).
We can summarize the characteristics of the observed trend as follows.

\subsubsection{Lines of Neutral Light Elements (Major Population)}
Regarding C~{\sc i} and O~{\sc i} lines of major population species, 
the $W$ values of high-excitation C~{\sc i} lines decrease (Figure~5a)
while those of low-excitation O~{\sc i} (forbidden) lines increase (Figure~5b) 
as $\cos\theta$ decreases from the disk center to the limb. 

\subsubsection{Lines of Neutral Heavier Elements (Minor Population)}
Generally, lines of neutral heavier ($Z>10$) species, which are of
minor population stage, are progressively strengthened toward the limb 
without exception (Figures 5c, 5d, 5e, 5f, 5h, 5i, 5m, 5n, 5q, 5r). 
The gradient of increase ($|{\rm d}\log W/{\rm d}\cos\theta|$) becomes larger
as the excitation potential ($\chi_{\rm low}$) decreases (Figures 5d,
5h, 5l, 5m, 5q). Moreover, this gradient tends to diminish as
the line becomes stronger (at the same $\chi_{\rm low}$; Figures 5e,
5f, 5i, 5n, 5r). We note that these trends were already reported
by Jevremovi\'{c} {\it et al.} (1993). 

\subsubsection{Lines of Ionized Species (Major Population)}

The behavior of $W$ in this group of lines is somewhat complicated.
That is, lines tend to be gradually intensified toward the limb at lower 
$\chi_{\rm low}$, while this trend is inverse at higher $\chi_{\rm low}$ 
(Figure~5o). The tendency of smaller gradient with increasing the 
line strength at the same $\chi_{\rm low}$ (as seen in neutral species) 
is also observed (Figures 5j, 5p, 5t). Generally, although many lines
of ionized species are strengthened (as we go from the disk center to the limb)
such like the case of neutral species, the rate of increase tends to be
rather smaller and near-flat cases are often seen (Figures~5k, 5l).

% Figure 6
\subsection{Physical Interpretation}

The next task is to give a reasonable interpretation on the diversified 
behaviors of $\log W$ {\it vs.} $\cos\theta$ relations revealed by various spectral
lines (Section~4.2). Here, the key is the sensitivity of line strength 
to a change in temperature, which differs from line to line.
We define the $T$-sensitivity of $W$ for each line by 
$K_{00} [\equiv ({\rm d}\log W/{\rm d}\log T)_{00}]$, 
which we numerically evaluated at the disk center (similarly to the procedure 
in TU17b) as follows:
\begin{equation}
K_{00} \equiv (W_{00}^{+100} - W_{00}^{-100})/W_{00}/(200/5780),
\end{equation}
where $W_{00}^{+100}$ and $W_{00}^{-100}$ are the equivalent widths computed (with 
the same $\log\epsilon$ solution used to derive the disk-center value of $W_{00}$) 
by two model atmospheres with $T_{\rm eff}$ perturbed by $+100$~K 
($T_{\rm eff} = 5880$~K and $\log g = 4.44$) and  $-100$~K
($T_{\rm eff} = 5680$~K and $\log g = 4.44$), respectively.
The resulting $K_{00}$ for each line is presented in Table~3.

In order to quantify the observed gradient of $\log W$ with a change of $\cos\theta$, 
we applied the linear-regression analysis to the set of 
($\cos\theta_{i}, \log W_{i}, i=00, \cdots, 31$)
to approximate $\log W$ with a linear function of $\cos\theta$ 
\begin{equation}
\log W = \alpha - \beta \cos\theta
\end{equation} 
and determined the parameters ($\alpha, \beta$) for each line, which are given in Table~3
(note that $\beta > 0$ for increasing $W$ toward the limb).
Such determined linear relations are also depicted by solid lines in each panel of Figure~5.

The resulting $\beta$ values are plotted against $\chi^{\rm ion} - \chi_{\rm low}$
(for minor population lines) or against $\chi_{\rm low}$ (for major population lines) 
in Figures~6a (all lines) and 6a$^{\prime}$ (Fe~{\sc i} and Fe~{\sc ii} lines). 
Further, in Figures~6b and 6b$^{\prime}$ are shown the distributions of $K_{00}$ in a similar manner. 
It is interesting to note that the both $\beta$ and $K_{00}$ exhibit quite similar dependence upon 
$\chi^{\rm ion} - \chi_{\rm low}$ or $\chi_{\rm low}$. This clearly indicates
that the center--limb variation of $W$ is mainly determined by the temperature sensitivity
specific to each spectral line, reflecting the fact that the line-forming layer progressively 
shifts toward upper/shallower layer of comparatively lower $T$ as we move from the disk center
to the limb.   

It is possible to explain this trend from the viewpoint of line-formation theory.
In the weak-line case where $W$ is almost proportional to the number population ($n$) 
of the lower level, the $\chi_{\rm low}$-dependence of $K$ is approximately expressed as
\begin{equation}
K^{\rm minor} \approx -11604 \, (\chi^{\rm ion} -\chi_{\rm low})/T \; (<0)
\end{equation} 
and 
\begin{equation}
K^{\rm major} \approx +11604 \, \chi_{\rm low}/T \; (>0)
\end{equation}
for minor- and major-population cases, respectively,
where  $\chi^{\rm ion}$ and $\chi_{\rm low}$ are in unit of eV and $T$ is in K 
({\it cf.} Section~4.1 in TU17b).
Actually, we can confirm from Figures~6b and 6b$^{\prime}$ that the trend of $K_{00}$ for
weak lines (small symbols, which constitute the upper and lower envelopes of the
distribution) roughly follow these two relations.
Then, as lines get stronger and more saturated, $W$ is not proportional to $n$
any more and its $T$-sensitivity ($|K_{00}|$) accordingly becomes smaller 
than that given by Equations~8 and 9, which explains why stronger lines 
tend to show progressively smaller $|K_{00}|$ values compared to weak lines 
in Figures~6b and 6b$^{\prime}$.

We point out that what has been mentioned above satisfactorily accounts for 
the observed trends of various $\log W$ {\it vs.} $\cos\theta$ relations summarized 
in Section~4.2; {\it i.e.}, (a) negative $\beta$ for high-excitation lines of 
major population species, (b) positive $\beta$ for the lines of minor population 
species, (c) $\chi_{\rm low}$-dependence of $\beta$, and (d) $|\beta|$ becoming 
smaller with increasing $W$ (when compared at the same $\chi_{\rm low}$). 
Accordingly, we can state that the observed center--limb variations of 
the strengths of solar spectral lines are reasonably understood in terms
of the properties of individual lines.\footnote{To be more complete,
it may be appropriate here to remark that, while the description given here
roughly explains the $\chi_{\rm low}$- or $W$-dependence of $\beta$
in the qualitative sense, it does not necessarily reproduce the absolute
value of $\beta$. For example, the strengths of low-excitation lines of 
major population species apparently increase toward the limb 
({\it e.g.}, O~{\sc i} 5577.339 or 6363.776 in Figure~5b, Sc~{\sc ii} 5552.224
in Figure~5g, Fe~{\sc ii} 6269.967 in Figure~5o; see the open symbols of 
low $\chi_{\rm low}$ in Figure~6a). This can not be explained by Equation~9,
which always yields positive $K$ and thus negative $\beta$. 
Actually, the $\chi_{\rm low}$-effect is not important for such low $\chi_{\rm low}$ 
lines of major population species, which are mainly controlled by the continuum 
opacity (the denominator of line-to-continuum opacity ratio which determines 
the line strength) because of the inert nature of the number population 
in the dominant ionization stage (see, {\it e.g.}, Section~2.1 in Takeda, Ohkubo, 
and Sadakane, 2002). Since the H$^{-}$ ion is the main source of continuum opacity 
which decreases as the density is lowered, we may interpret that the increase 
of $W$ toward the limb in this case is due to a decrease of density caused 
by an upward shift of the line-forming layer.}

% Figure 7
\subsection{Comparison with Published Studies}

As already mentioned in Section~1, only a small number of investigations 
have been carried out so far on the center-to-limb variations of spectral line strengths.
Moreover, most of the such studies are rather outdated (done several decades
ago) and few recent work based on the modern technique is available.   
In any event, it may be worthwhile to compare our $W$ {\it vs.} $\cos\theta$ results
with those of published studies. 

Such comparisons for representative 20 lines are depicted in Figure~7, 
which reveals that the consistency is not necessarily satisfactory. 
Although there are several lines for which good agreement is seen ({\it e.g.}, 
Figures~7h, 7t), appreciable discrepancies are observed in quite a few cases.
Above all, comparisons with Balthasar's (1988) measurements (in which systematic 
errors are suspected as mentioned in Section~3.3) show appreciable disagreement. 
Still, if we disregard the matching of absolute $W$ values, we may state that 
the global trends of center--limb variation ({\it i,e.}, whether increasing or decreasing) 
are more or less similar. 

Yet, there are some cases where even the sense of variation is opposite. 
For example, regarding Fe~{\sc i} 5930.173,  Rodr\'{\i}guez Hidalgo,
Collados, and V\'{a}zquez (1994) reported a decreasing tendency of $W$ 
toward the limb ({\it cf.} their Fig.~2b), which is in conflict with our 
results (Figure~7j) and hard to explain for such Fe~{\sc i} lines 
of minor population stage.
Likewise, we notice in Figure~7l that the equivalent width of 
Fe~{\sc i} 6093.666 derived by the Soviet group 
(Gurtovenko, and Ratnikova, 1976; Kostik, 1982) slightly decreases 
(in contrast with our result which slightly increases) toward the limb. 
However, since the $\mu$-dependence of $W$ is rather small and near-flat 
for such a Fe~{\sc i} line of comparatively higher excitation 
($\chi_{\rm low} = 4.61$~eV, $\beta = +0.103$), this disagreement 
may be regarded as within tolerable uncertainties (actually, 
the $W$ values themselves are favorably compared with each other;
{\it cf.} Figure~7l).

%Section 5. Conclusion
\section{Conclusion}

Observational studies have been rather insufficient regarding the solar center-to-limb 
variations of spectral line strengths, despite that they contain valuable information 
on the physical structure of the photosphere. Actually, available are only several 
investigations mostly done a few decades ago, which are outdated from the viewpoint 
of present-day standard. 

Recently, we found that application of our semi-automatic synthetic 
spectrum-fitting technique to the profiles of solar spectral lines is 
an effective method, by which the equivalent widths of spectral lines 
can be evaluated at a number of points on the solar disk quite efficiently 
(TU17a, TU17b). 

Accordingly, equipped with this technique, we decided to conduct a comprehensive study 
on the solar center--limb variations of equivalent widths ($W$) for a large number 
of spectral lines based on the observational data obtained by the Domeless Solar 
Telescope at Hida Observatory, in order to clarify the behaviors of line strengths 
across the solar disk and to understand them in terms of line properties.

As such, the $W$ values of 565 selected spectral lines in the wavelength range
of 4690--6870~\AA\ were evaluated at 31 consecutive points from the solar disk 
center ($\mu = 1$) near to the limb ($\mu = 0.25$)
In order to quantify the global variation of $W$ with a change of $\mu$, 
the gradient $\beta (\equiv - {\rm d}\log W/{\rm d}\log\mu)$ was determined for each line 
by the linear-regression analysis.

It turned out that most lines are strengthened ($\beta > 0$) while a small fraction 
of lines are weakened ($\beta < 0$) toward the limb, and that the values of
$\beta$ depend on the excitation potential as well as line strengths.
Interestingly, the distribution of $\beta$ was found to well correlates with 
that of $K (\equiv {\rm d}\log W/{\rm d}\log T)$, which means that the center-to-limb variation 
of $W$ is mainly controlled by the $T$-sensitivity of individual lines because 
the line-forming region shifts towards upper layers of lower $T$ as we approach the limb. 

Further, it was shown that more physical insight can be gained by dividing 
the lines into two groups, minor population stage (most neutral species of $Z>10$) 
or major population stage (neutral light species such as C~{\sc i} or O~{\sc i}
and all ionized  species), by which the $\chi_{\rm low}$-dependent trend of $K$
could be reasonably explained. 
 
All the center--limb data of the equivalent widths (as well as the line-of-sight
turbulent velocity dispersions, elemental abundances, and mean line-formation depths 
derived as by-products) along with the solar spectra used for 
our analysis are available as on-line materials.   
\newline
\newline
{\bf Disclosure of Potential Conflicts of Interest}\\
The authors declare that they have no conflicts of interest.

\appendix
%Appendix A. (Table 4, Table 5)
\section{Supplementary Materials}

We present the full results of our analysis (along with the used spectra)
as on-line materials, which consist of three parts as described below.

% Table 4
\subsection{Atomic Line Data and Summary of the Results}

This is a data table (``tableE.dat'': a text file of 565 lines of 175 bytes) 
including the basic atomic data and the brief results of our analysis,
which is an extension of Table~3. See Table~4 for the details
regarding the contents and the data format of this table.

\subsection{Solar Spectra Used in This Study}

Our observations for a given spectral range were done at 32 targeted points 
on the solar disk (00, 01, 02, $\cdots$, 30, 31; {\it cf.} Table~1), which were 
repeated for 109 wavelength regions (each covering $\approx$~22--25~\AA,  
and named as W4700, W4720, W4740, $\cdots$, W6840, and W6860; {\it cf.} Table~2).
All spectra are combined and presented in a single large file ``all\_spectra'',
from which each of the 3488 ($= 32 \times 109$) individual spectra 
``w????rad??.txt'' can be extracted by using the fortran program ``spec\_extract.for''.
For example, ``w5200rad00.txt'' is the 5200\AA-centered spectrum at point 00 
(disk center), and ``w6780rad31.txt'' is the 6780\AA-centered spectrum at 
point 31 (nearest to the limb). 
Each spectrum file comprises 1600 lines (corresponding to 1600 wavelength points
with a step of 0.015~\AA) of 18 bytes; and the wavelength (in \AA) and
the residual intensity (normalized by the local continuum) are given
in the format of (F9.3, F9.5) at each line. 

Our spectra should not be used for any purpose requiring high wavelength 
precision ({\it e.g.}, study of limb effect, {\it etc.}), because their wavelength
scales are not so accurately calibrated (which were done only approximately 
by comparing the solar spectral lines with the published solar FTS spectrum atlas;
{\it cf.} Section~2). The precision of the wavelength scale can be guessed 
by examining how the spectra of overlapping parts (of several \AA) of two 
consecutive regions match each other. 
The mutual differences of wavelength scale for 108 overlapping portions (of 
109 regions in Table~2) determined applying the cross-correlation technique 
($\Delta \lambda_{j} \equiv \lambda_{j+1} - \lambda_{j}$; $j = 1, 2, \cdots, 108$) 
turned out to range from $-0.032$~\AA\ to $+0.024$~\AA\, almost following the
normal distribution (around the mean $\langle \Delta \lambda \rangle$ of 
$-0.002$~\AA) with the standard deviation of $\pm 0.010$~\AA. Accordingly, 
we may state that our wavelength precision is typically $\approx$~0.01-0.02~\AA\
(corresponding to $\approx 1$~km~s$^{-1}$ in the velocity scale).

% Table 5
\subsection{Center--Limb Variations of Observed Quantities}

All the center-to-limb data of the equivalent widths, abundances, 
mean formation depths, and line-of-sight turbulent velocity dispersions
derived from our spectrum fitting analysis for 565 lines are combined 
and presented in a single large file ``all\_clvdata'', from which the individual 
data file ``????\_????????.dat'' for each line can be extracted by using the 
fortran program ``clv\_extract.for''.
The 12-character string ``????\_????????'' is the spectral line code
constructed from the species code and the wavelength as in tableE.dat ({\it cf.} Table~4).
The first line of each file is the header including the basic line data as well as
the results of linear-regression coefficients (as in tableE.dat):
[s-code, $\lambda$, $\chi_{\rm low}$, $\log gf$, $\alpha$, $\beta$ 
(1x,F6.2,F9.3,F7.3,F8.3,F9.3,F7.3)].
In the following 32 lines are given the data at each of the 32 points 
(00, 01, 02, ..., 30, 31). See Table~5 regarding the contents and their data format.
Note that dummy values ({\it e.g.}, $-9.99$ or $-9.999$) are given for the indeterminable 
case ({\it i.e.}, where satisfactory convergence was not accomplished).

\newpage
%% Figure 1
\begin{figure} 
\centerline{\includegraphics[width=0.75\textwidth]{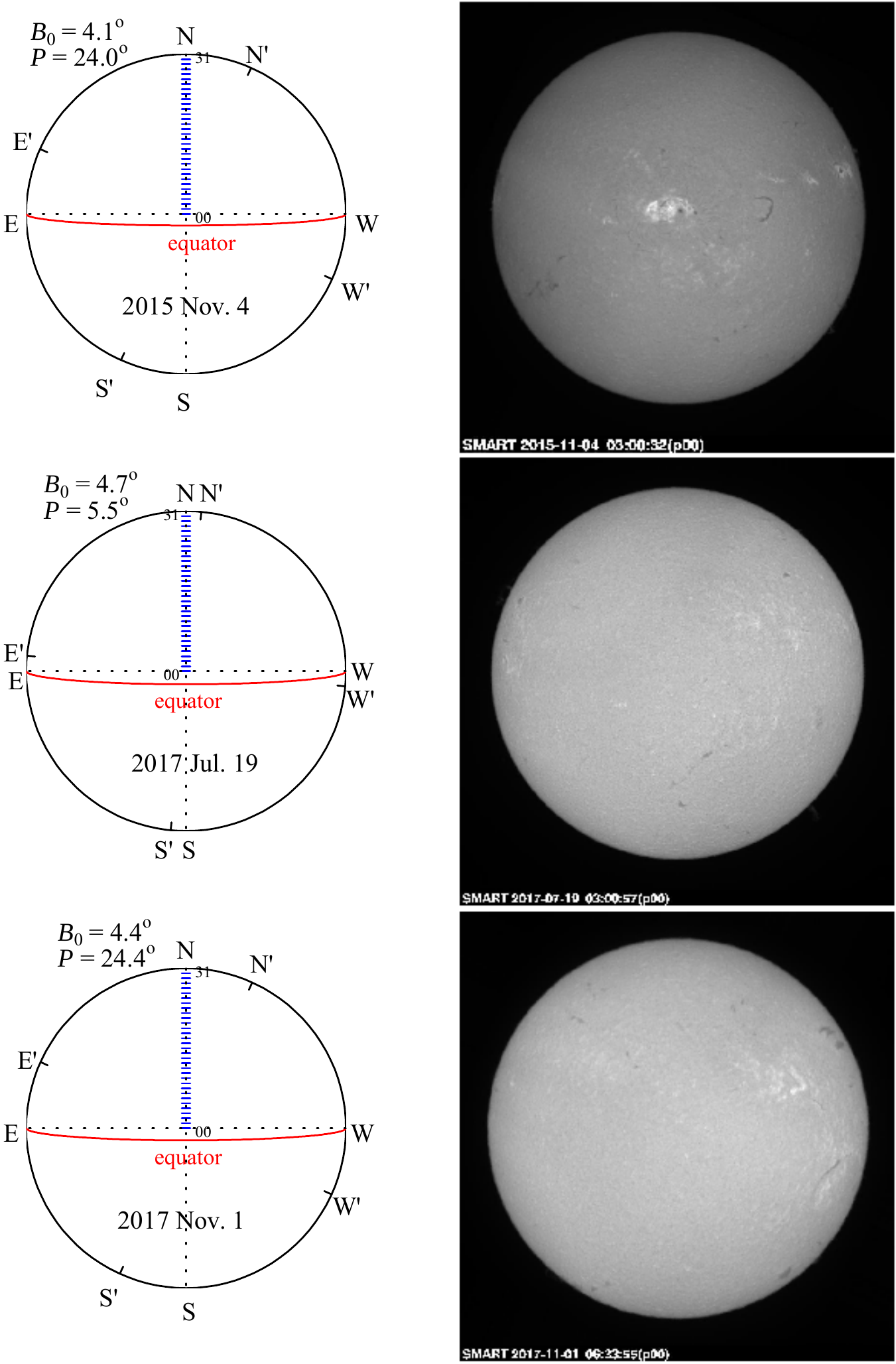}}
\caption{
Left panels: Graphical description of the observed points on the solar disk,
(32 points on the northern meridian from the disk center 
to $0.97 {\rm R}_{\odot}$ with a step of $30'' \approx 0.03{\rm R}_{\odot}$, while
spatially averaged over $50''$ along the E--W direction)
at which the spectral data obtained in this study were taken.
Shown are the three configurations of the Sun on the mid-dates of three 
different observing seasons (2015 November, 2017 July, and 2017 November).  
N, S, E, and W are the directions in reference to the Sun (based on 
solar rotation), whereas those in the equatorial coordinate system on 
the celestial sphere (defined by the rotation of Earth) are also denoted 
as N$'$, S$'$, E$'$, and W$'$.
Right panels: Corresponding H$\alpha$ full-disk images of the Sun on these dates, 
which were observed by Solar Magnetic Activity Research Telescope (SMART) 
at Hida observatory.  
}%\label{}
\end{figure}

%% Figure 2
\begin{figure} 
\centerline{\includegraphics[width=0.75\textwidth]{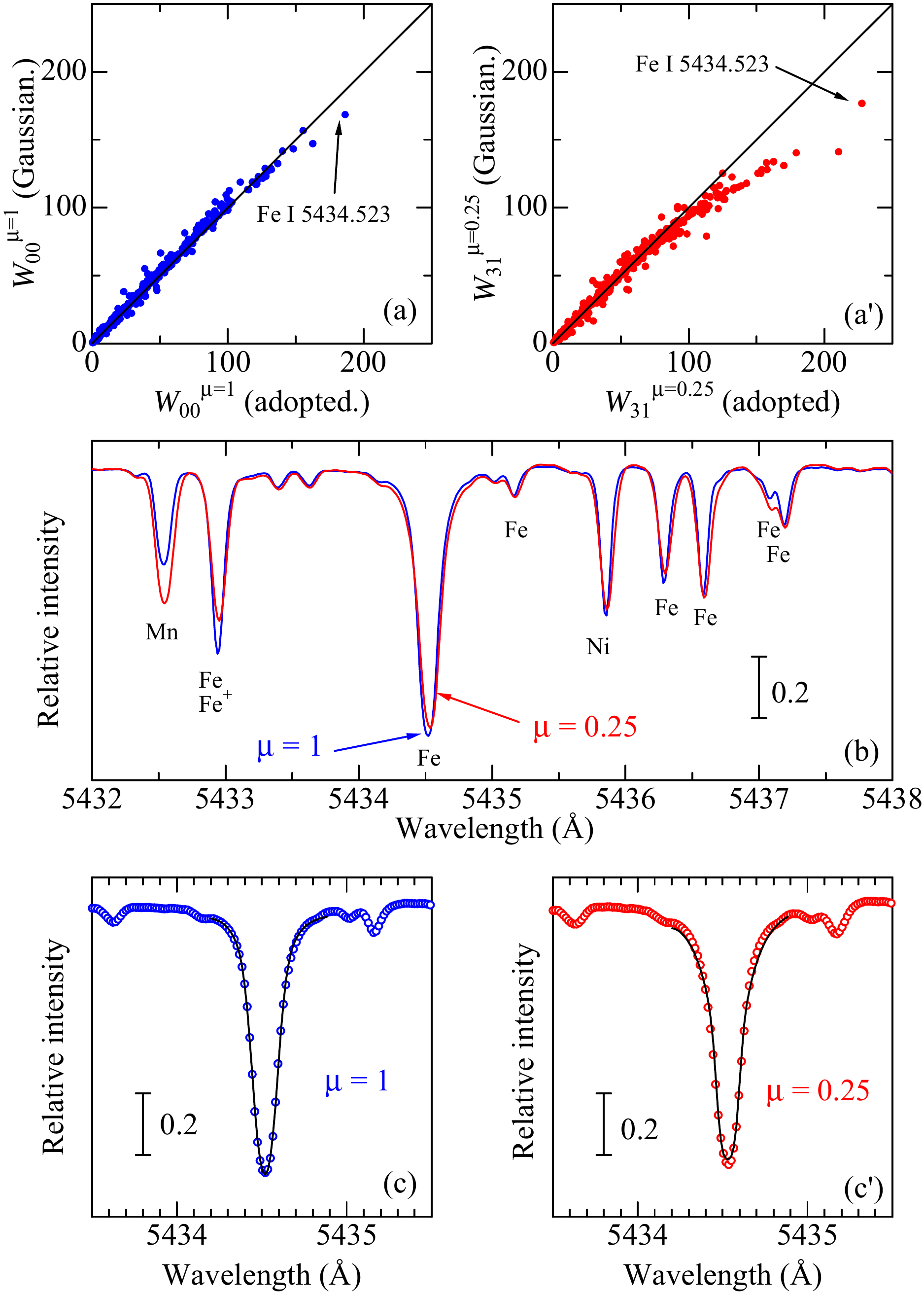}}
\caption{
Upper panels (a) and (a$^{\prime}$): comparison of the finally obtained equivalent 
widths (abscissa) with those measured by the Gaussian-fitting technique 
(ordinate), where left (a) and right (a$^{\prime}$) panels are for the disk center 
(point~00; $\mu = 1$) and nearest to the limb (point~31; $\mu = 0.25$),
respectively. 
Middle panel (b): Comparison of disk-center spectrum ($\mu=1$: blue line) 
with the nearest-to-limb one ($\mu=0.25$: red line) in the 
5432--5438~\AA\ region. 
Lower panels (c) and (c$^{\prime}$):  Comparison of the theoretically modeled profile 
(lines) of Fe~{\sc i} 5434.523, which is best-fitted with the observed one 
(symbols) for $\mu=1$ (left) and $\mu=0.25$ (right) cases. 
}%\label{}
\end{figure}

%% Figure 3
\begin{figure} 
\centerline{\includegraphics[width=0.8\textwidth]{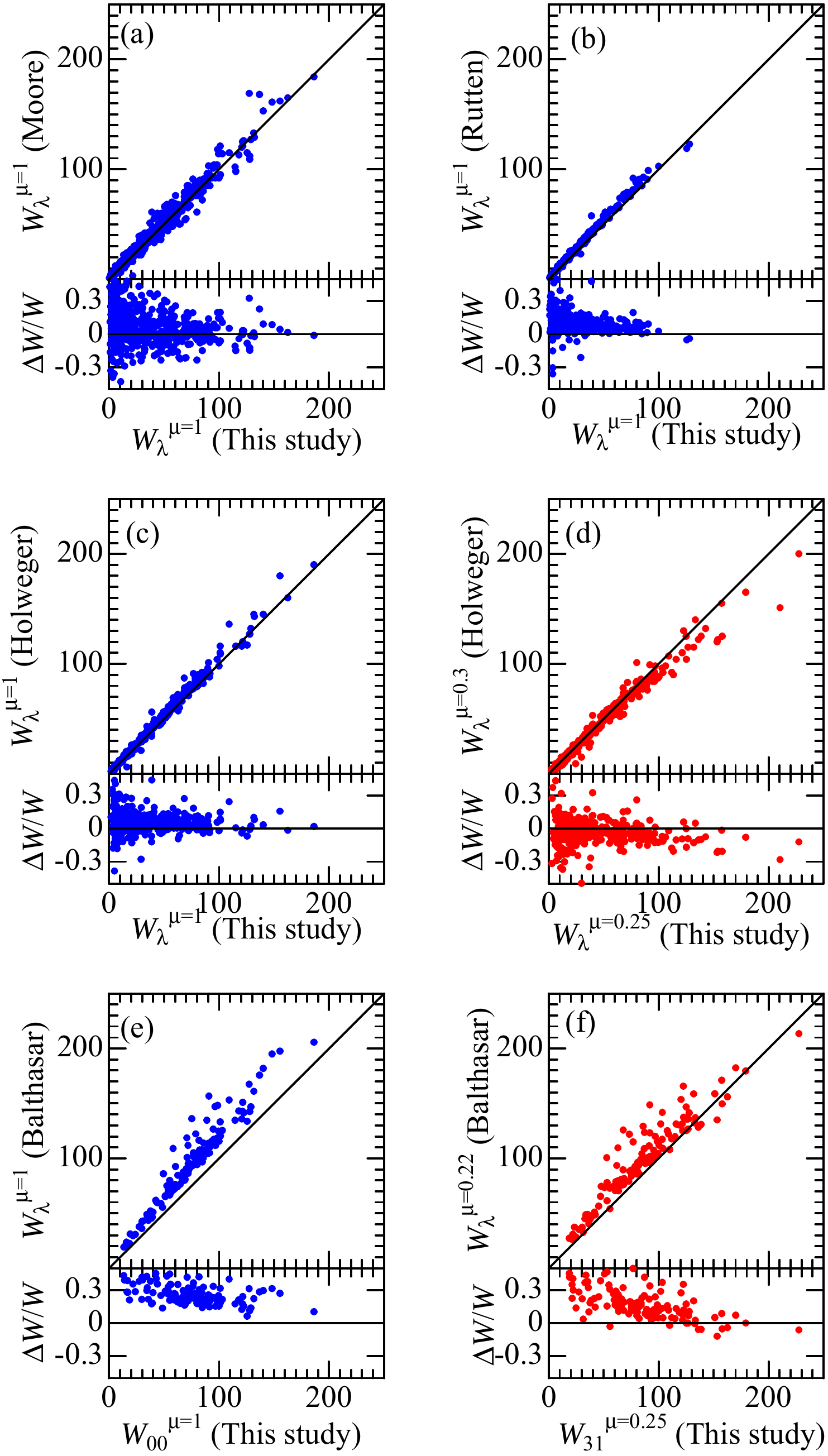}}
\caption{
Our equivalent widths are compared with those taken from four representative 
publications, where not only the direct comparison 
[$W_{\rm other}$ {\it vs.} $W_{\rm our}$; upper panel] but also the behaviors of relative 
differences [$(W_{\rm other}-W_{\rm our})/W_{\rm our}$ {\it vs.} $W_{\rm our}$; lower panel] 
are shown.
The figure panels with blue symbols are for $\mu=1$ (disk center), 
while those with red symbols are for $\mu =$~0.2--0.3 (near the limb).
(a) Moore, Minnaert, and Houtgast (1966). (b) Rutten and van der Zalm (1984a)
(their background-corrected $W^{\rm T}$ values were used).
(c),(d) Holweger (1967). (e),(f) Balthasar (1988).
}%\label{}
\end{figure}

%% Figure 4
\begin{figure} 
\centerline{\includegraphics[width=0.7\textwidth]{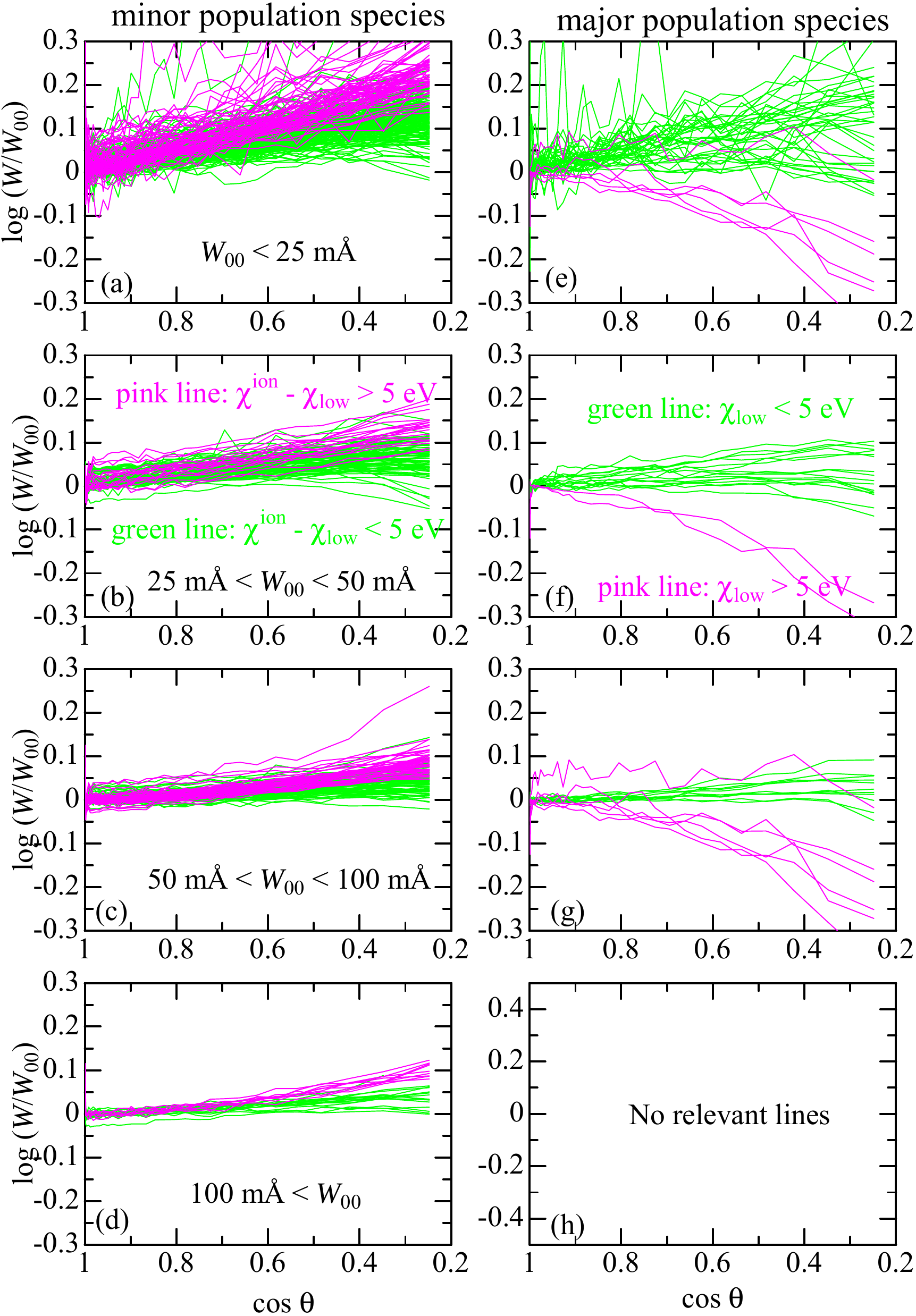}}
\caption{
Behaviors of $\log(W/W_{00})$ (equivalent width ratio normalized by 
the disk-center value) plotted against $\cos\theta$.
The left panels (a)--(d) and the right panels (e)--(h) are for the lines
of minor population stage ({\it e.g.}, Fe~{\sc i}) and those of major population
stage ({\it e.g.}, C~{\sc i}, Fe~{\sc ii}), respectively.
Lines are divided into four groups according to the range of $W_{00}$:
$W_{00} < 25$~m\AA\ (a,e),  25~m\AA~$\le W_{00} < 50$~m\AA\ (b,f),
 50~m\AA~$\le W_{00} < 100$~m\AA\ (c,g), and 100~m\AA~$\le W_{00}$ (d,h).
In each panel, lines are also discriminated into two classes and plotted 
in different colors according to the relevant key potential energy:
For left panels (a--d: minor population),
$\chi^{\rm ion} - \chi_{\rm low} < 5$~eV (green) and
$\chi^{\rm ion} - \chi_{\rm low} > 5$~eV (pink).
For right panels (e--h: major population),
$\chi_{\rm low} < 5$~eV (green) and 
$\chi_{\rm low} > 5$~eV (pink).
}%\label{}
\end{figure}

\clearpage
%% Figure 5
\begin{figure} 
\centerline{\includegraphics[width=1.\textwidth]{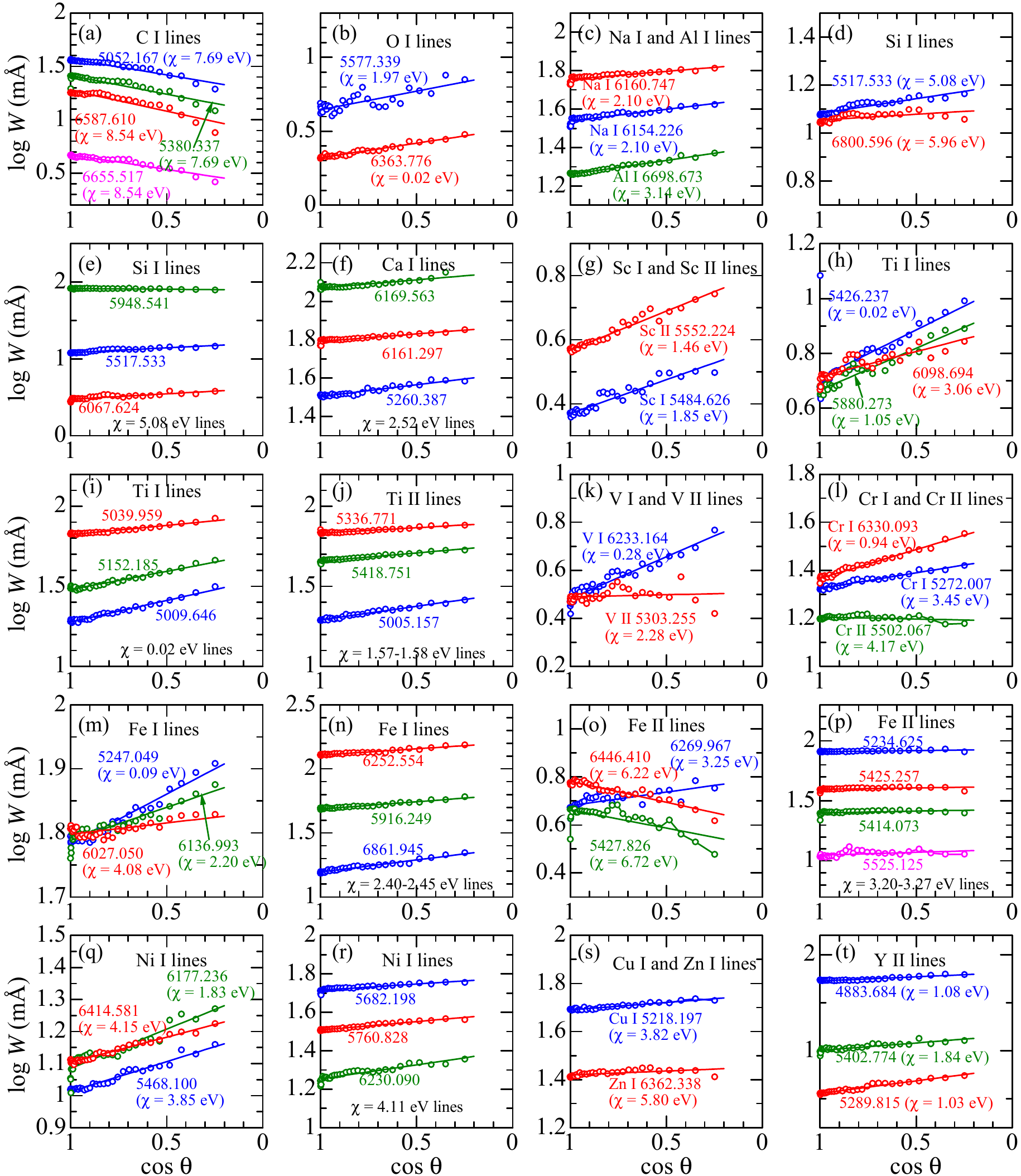}}
\caption{
The $\log W$ values of various representative lines are plotted 
against $\cos\theta$ in order to demonstrate their center--limb variations. 
The spectral lines displayed in each panel are appropriately selected 
in order to clarify the characteristic trends in terms of the species, 
excitation potentials, and line strengths. The observed data are shown by open circles 
while the corresponding linear-regression lines ({\it cf.} Equation~7) are depicted
by solid lines.
}%\label{}
\end{figure}

%% Figure 6
\begin{figure} 
\centerline{\includegraphics[width=0.75\textwidth]{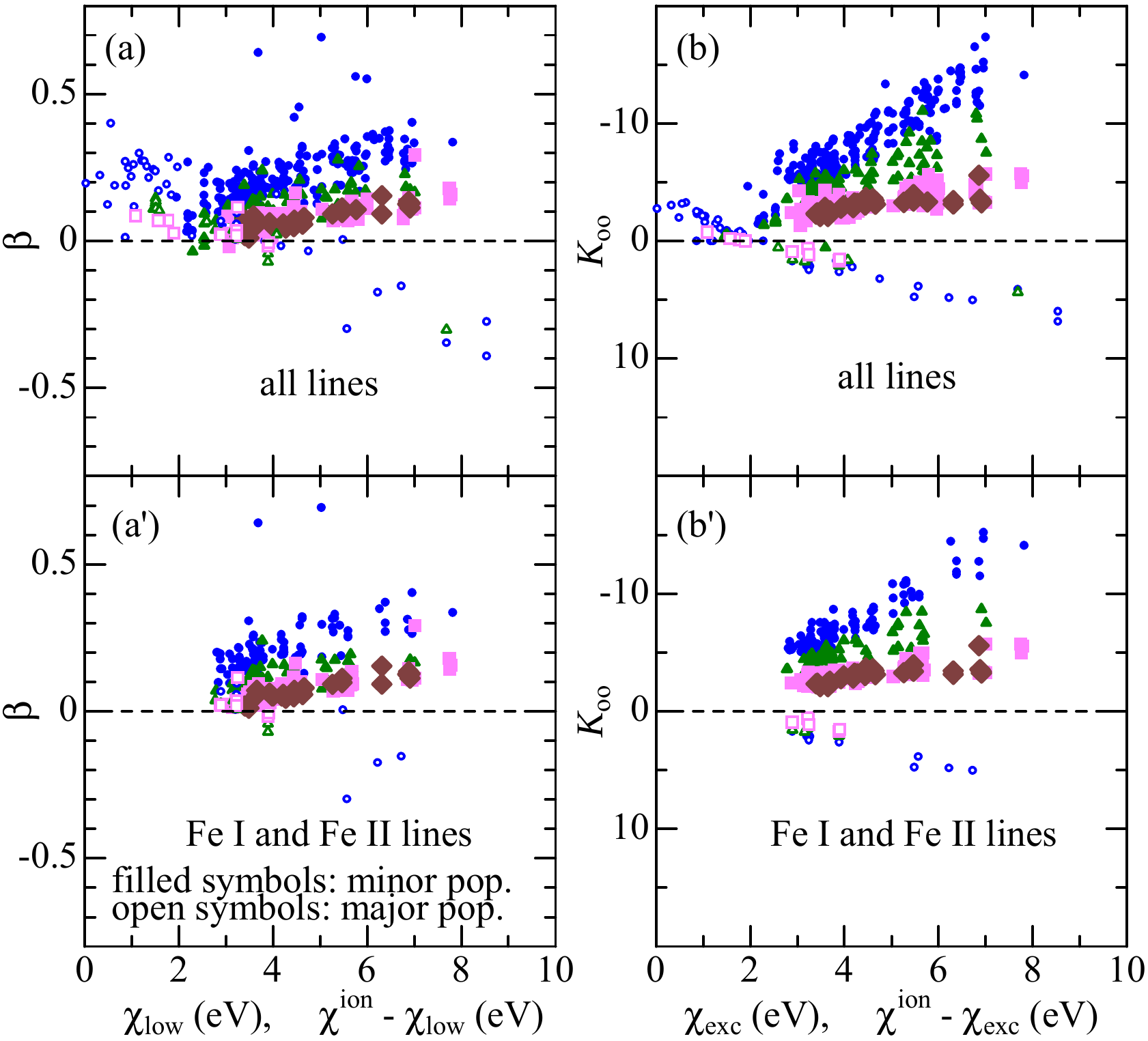}}
\caption{
The behaviors of $\beta$ (left panels (a) and (a$^{\prime}$)) and $K_{00}$ 
(right panels (b) and (b$^{\prime}$)) 
plotted against $\chi_{\rm low}$ (for major population species) or 
$\chi^{\rm ion} - \chi_{\rm low}$ (for minor population species).
The upper panels (a) and (b) show the results of all lines,
while the lower panels (a$^{\prime}$) and (b$^{\prime}$) display those 
for Fe~{\sc i} and Fe~{\sc ii} lines. 
The filled and open symbols correspond to minor population
and major population species, respectively.
Lines of different strengths classes are discriminated by 
the shape and the size (larger for stronger lines) of symbols:
circles (blue): $W_{00} <$~25~m\AA\, 
triangles (green): 25~m\AA~$\le W_{00} <$~50~m\AA\,
squares (pink): 50~m\AA~$\le W_{00} <$~100~m\AA\, and
diamonds (brown): 100~m\AA~$\le W_{00}$.
}%\label{}
\end{figure}

%% Figure 7
\begin{figure} 
\centerline{\includegraphics[width=1.\textwidth]{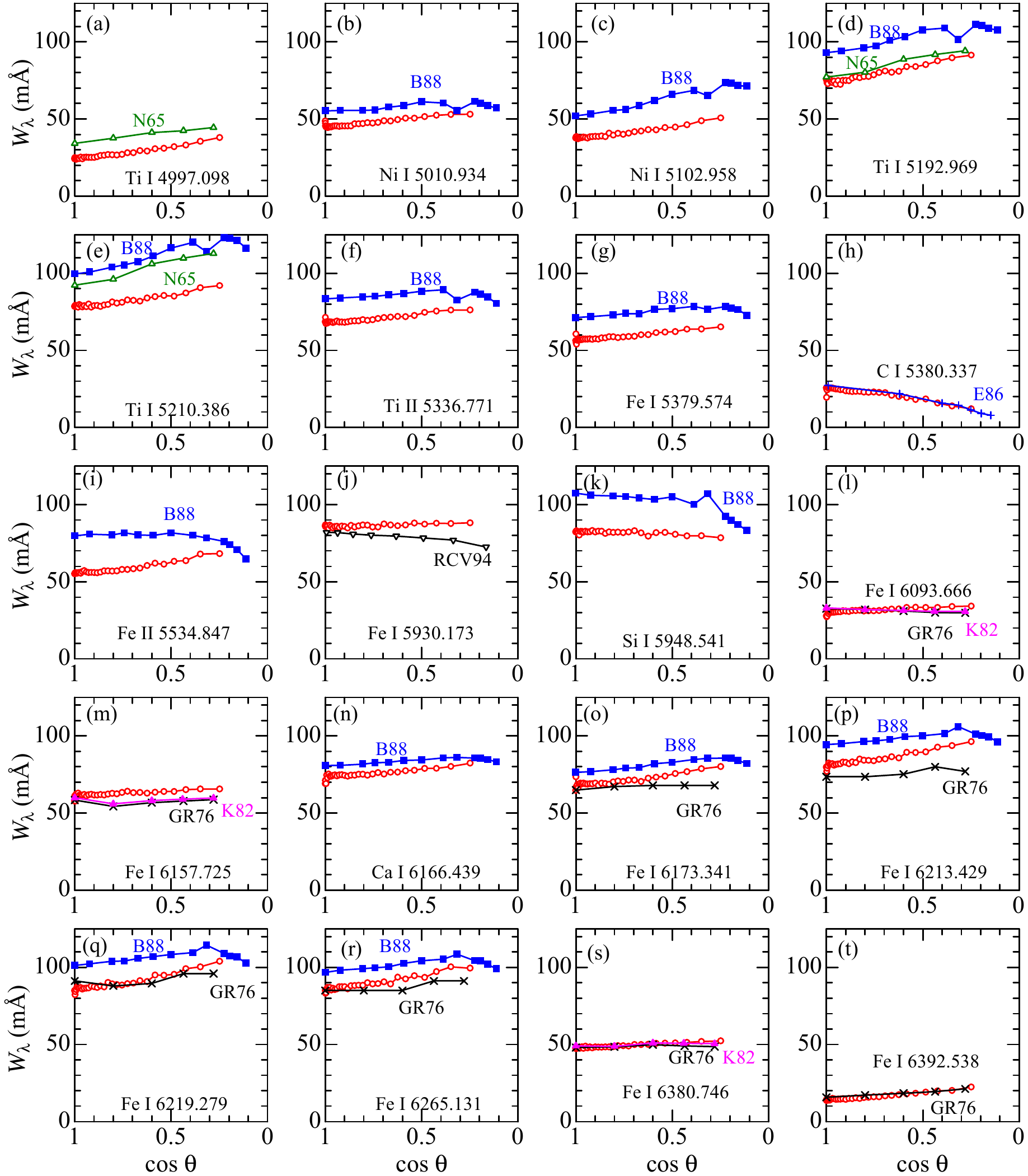}}
\caption{
In each panel are compared our $W$ {\it vs.} $\cos\theta$ relations derived for 
20 lines (open circles) with those published by previous studies. 
N65 (green open triangles): Nissen (1965), GR76 (black St Andrew's crosses): 
Gurtovenko and Ratnikova (1976), K82 (pink filled triangles): Kostik (1982), 
E86 (blue Greek crosses): Elste (1986), B88 (blue filled squares): Balthasar (1988), 
and RCV94 (black open inverse triangles): Rodr\'{\i}guez Hidalgo, Collados, 
and V\'{a}zquez (1994).
}%\label{}
\end{figure}

\newpage

%Table 1
%\clearpage
\setcounter{table}{0}
\begin{table}[h]
%\tiny
\scriptsize
%\small
\caption{Observed points on the solar disk.}
\begin{center}
\begin{tabular}
{cccc c cccc}\hline\hline
No.& $\sin\theta$ & $\cos\theta$ & $\theta$ &  & No.& $\sin\theta$ & $\cos\theta$ & $\theta$\\
(1) & (2) & (3) & (4) &   & (1) & (2) & (3) & (4) \\
\hline
00 & 0.000  & 1.000 & 0.0  &  & 16 & 0.500  & 0.866 & 30.0 \\
01 & 0.031  & 1.000 & 1.8  &  & 17 & 0.531  & 0.847 & 32.1 \\
02 & 0.063  & 0.998 & 3.6  &  & 18 & 0.563  & 0.827 & 34.2 \\
03 & 0.094  & 0.996 & 5.4  &  & 19 & 0.594  & 0.805 & 36.4 \\
04 & 0.125  & 0.992 & 7.2  &  & 20 & 0.625  & 0.781 & 38.7 \\
05 & 0.156  & 0.988 & 9.0  &  & 21 & 0.656  & 0.755 & 41.0 \\
06 & 0.188  & 0.982 & 10.8 &  & 22 & 0.688  & 0.726 & 43.4 \\
07 & 0.219  & 0.976 & 12.6 &  & 23 & 0.719  & 0.695 & 46.0 \\
08 & 0.250  & 0.968 & 14.5 &  & 24 & 0.750  & 0.661 & 48.6 \\
09 & 0.281  & 0.960 & 16.3 &  & 25 & 0.781  & 0.624 & 51.4 \\
10 & 0.313  & 0.950 & 18.2 &  & 26 & 0.813  & 0.583 & 54.3 \\
11 & 0.344  & 0.939 & 20.1 &  & 27 & 0.844  & 0.537 & 57.5 \\
12 & 0.375  & 0.927 & 22.0 &  & 28 & 0.875  & 0.484 & 61.0 \\
13 & 0.406  & 0.914 & 24.0 &  & 29 & 0.906  & 0.423 & 65.0 \\
14 & 0.438  & 0.899 & 25.9 &  & 30 & 0.938  & 0.348 & 69.6 \\
15 & 0.469  & 0.883 & 28.0 &  & 31 & 0.969  & 0.248 & 75.6 \\
\hline
\end{tabular}
\end{center}
%\small
\scriptsize
%\tiny
(1) Designated number of each observed point ({\it cf.} Figure~1). (2) Value of $\sin\theta$ (equivalent to 
the concentric radius in unit of the solar radius), where $\theta$ is the angle between 
the line of sight and the normal to the surface. (3) Value of $\cos\theta$, which is often 
referred to as $\mu$. (4)  Value of $\theta$ (in deg). 
Note that all the values given here correspond to the positions just on the meridian.
Since our spectra were derived by spatially averaging along the direction
of the slit (perpendicular to the meridian) by $\approx \pm 25''$ ({\it cf.} Section~2), 
light from slightly different (larger) angle is actually included: 
{\it i.e.}, up to $\theta' (> \theta)$ where 
$\sin \theta' \equiv \sqrt{\sin^{2} \theta + (25/960)^{2}}$.
Nevertheless, this effect is insignificant, which is barely detectable 
only near the disk center of small $\theta$; {\it e.g.}, 
$\sin \theta' - \sin \theta$ is 0.026 (No.~00), 0.010 (No.~01), 
0.005 (No.~02), 0.002 (No.~05), and 0.001 (No.~10).   
\end{table}

%Table 2
%\clearpage
\setcounter{table}{1}
\begin{table}[h]
\tiny
%\scriptsize
%\small
\caption{Wavelength regions and observed dates of the spectra.}
\begin{center}
\begin{tabular}
{cccc c cccc}\hline\hline
Region& $\lambda_{\rm min}$ & $\lambda_{\rm max}$ & Date &  & Region& $\lambda_{\rm min}$ & $\lambda_{\rm max}$ & Date \\
(1) & (2) & (3) & (4) &   & (1) & (2) & (3) & (4) \\
\hline
w4700  &  4687.5 & 4712.3 & 171103& & w5800  &  5788.8 & 5811.1 & 170720 \\
w4720  &  4707.6 & 4732.4 & 171103& & w5820  &  5808.7 & 5831.0 & 170720 \\
w4740  &  4727.7 & 4752.5 & 171103& & w5840  &  5827.7 & 5851.7 & 151104 \\
w4760  &  4747.7 & 4772.4 & 171103& & w5860  &  5847.9 & 5871.9 & 151104 \\
w4780  &  4767.5 & 4792.3 & 171103& & w5880  &  5868.1 & 5892.1 & 171101 \\
w4800  &  4787.6 & 4812.4 & 171103& & w5900  &  5888.0 & 5912.0 & 171101 \\
\hline
\end{tabular}
\end{center}
%\small
%\scriptsize
\tiny
(1) Region name. (2) Minimum wavelength (in \AA). (3) Maximum wavelength (in \AA).
(4) Observed date expressed as $yymmdd$ ({\it e.g.}, ``171103'' means that the spectrum
was observed on 2017 November 3).
\end{table}

%Table 3 [1/10]
\clearpage
\setcounter{table}{2}
\begin{table}[h]
\tiny
%\scriptsize
%\small
\caption{Data of the spectral lines and strength-related parameters. [1/10]}
\begin{center}
\begin{tabular}
{cccccrrrcc}\hline\hline
s-code& $\lambda$ & $\chi_{\rm low}$ & $\log gf$& $\log\epsilon_{00}$ &$W_{00}$ & $W_{31}$ & $K_{00}$ & $\alpha$ & $\beta$ \\ 
(1) & (2) & (3) & (4) & (5) & (6) & (7) & (8) & (9) & (10) \\
\hline
\hline
\multicolumn{10}{c}{[C~{\sc i} lines, $\chi^{\rm ion}$ =11.26~eV, major population stage]} \\
  6.00& 5052.167&  7.68&  $-$1.65&  8.73&  35.8&  19.3&  +4.36&  1.269& $-$0.302\\
  6.00& 5380.337&  7.68&  $-$1.84&  8.71&  25.7&  12.1&  +4.11&  1.067& $-$0.346\\
  6.00& 6587.610&  8.54&  $-$1.60&  8.97&  17.9&   7.6&  +5.99&  0.884& $-$0.391\\
  6.00& 6655.517&  8.54&  $-$1.37&  8.06&   4.7&   2.6&  +6.84&  0.398& $-$0.274\\
\hline
\multicolumn{10}{c}{[O~{\sc i} lines, $\chi^{\rm ion}$ =13.62~eV, major population stage]} \\
\hline
\end{tabular}
\end{center}
$^{a}$Partially blended with C$_{2}$ lines ({\it cf.} Section~3.2).\\
$^{b}$Partially blended with CN line ({\it cf.} Section~3.2).\\ 
\end{table}

%Table 3 [2/10]
%\clearpage
\setcounter{table}{2}
\begin{table}[h]
\tiny
%\scriptsize
%\small
\caption{(Continued.) [2/10]}
\begin{center}
\begin{tabular}
{cccccrrrcc}\hline\hline
s-code& $\lambda$ & $\chi_{\rm low}$ & $\log gf$& $\log\epsilon_{00}$ &$W_{00}$ & $W_{31}$ & $K_{00}$ & $\alpha$ & $\beta$ \\ 
(1) & (2) & (3) & (4) & (5) & (6) & (7) & (8) & (9) & (10) \\
\hline
 22.00& 4926.147&  0.82&  $-$2.17&  4.98&   5.3&  10.3& $-$12.84&  1.072& +0.355\\
 22.00& 4997.098&  0.00&  $-$2.12&  4.87&  24.1&  37.9& $-$11.80&  1.627& +0.245\\
 22.00& 4999.504&  0.83&  +0.25&  4.84&  95.3& 118.6&  $-$3.21&  2.088& +0.110\\
 22.00& 5009.646&  0.02&  $-$2.26&  4.90&  19.0&  31.5& $-$12.63&  1.546& +0.268\\
 22.00& 5016.162&  0.85&  $-$0.57&  4.87&  58.6&  67.7&  $-$5.17&  1.853& +0.112\\
 22.00& 5022.871&  0.83&  $-$0.43&  4.92&  68.7&  80.8&  $-$4.21&  1.930& +0.117\\
\hline
\end{tabular}
\end{center}
\end{table}

%Table 3 [3/10]
%\clearpage
\setcounter{table}{2}
\begin{table}[h]
\tiny
%\scriptsize
%\small
\caption{(Continued.) [3/10]}
\begin{center}
\begin{tabular}
{cccccrrrcc}\hline\hline
s-code& $\lambda$ & $\chi_{\rm low}$ & $\log gf$& $\log\epsilon_{00}$ &$W_{00}$ & $W_{31}$ & $K_{00}$ & $\alpha$ & $\beta$ \\ 
(1) & (2) & (3) & (4) & (5) & (6) & (7) & (8) & (9) & (10) \\
\hline
 23.00& 5670.853&  1.08&  $-$0.42&  3.96&  16.1&  24.4& $-$11.07&  1.427& +0.199\\
 23.00& 6058.139&  1.04&  $-$1.37&  4.13&   3.5&   5.7& $-$13.46&  0.803& +0.232\\
 23.00& 6111.645&  1.04&  $-$0.71&  3.90&   9.2&  14.1& $-$12.54&  1.236& +0.239\\
 23.00& 6135.361&  1.05&  $-$0.75&  3.85&   7.6&  13.3& $-$12.60&  1.198& +0.274\\
 23.00& 6150.157&  0.30&  $-$1.78&  4.01&   5.7&  11.3& $-$14.33&  1.148& +0.323\\
 23.00& 6224.529&  0.29&  $-$2.01&  4.06&   4.0&   6.8& $-$14.62&  0.909& +0.305\\
\hline
\end{tabular}
\end{center}
\end{table}

%Table 3 [4/10]
%\clearpage
\setcounter{table}{2}
\begin{table}[h]
\tiny
%\scriptsize
%\small
\caption{(Continued.) [4/10]}
\begin{center}
\begin{tabular}
{cccccrrrcc}\hline\hline
s-code& $\lambda$ & $\chi_{\rm low}$ & $\log gf$& $\log\epsilon_{00}$ &$W_{00}$ & $W_{31}$ & $K_{00}$ & $\alpha$ & $\beta$ \\ 
(1) & (2) & (3) & (4) & (5) & (6) & (7) & (8) & (9) & (10) \\
\hline
 26.00& 4813.116&  3.27&  $-$2.89&  7.55&  16.2&  22.5&  $-$7.80&  1.395& +0.195\\
 26.00& 4835.869&  4.10&  $-$1.50&  7.63&  45.0&  54.1&  $-$3.79&  1.766& +0.114\\
 26.00& 4882.144&  3.42&  $-$1.64&  7.67&  71.4&  81.9&  $-$2.91&  1.925& +0.079\\
 26.00& 4907.733&  3.43&  $-$1.84&  7.60&  58.8&  73.4&  $-$3.39&  1.879& +0.116\\
 26.00& 4939.686&  0.86&  $-$3.34&  7.38&  91.0& 112.2&  $-$3.27&  2.069& +0.113\\
 26.00& 4950.104&  3.42&  $-$1.67&  7.68&  71.5&  83.4&  $-$2.91&  1.943& +0.093\\
\hline
\end{tabular}
\end{center}
\end{table}

%Table 3 [5/10]
%\clearpage
\setcounter{table}{2}
\begin{table}[h]
\tiny
%\scriptsize
%\small
\caption{(Continued.) [5/10]}
\begin{center}
\begin{tabular}
{cccccrrrcc}\hline\hline
s-code& $\lambda$ & $\chi_{\rm low}$ & $\log gf$& $\log\epsilon_{00}$ &$W_{00}$ & $W_{31}$ & $K_{00}$ & $\alpha$ & $\beta$ \\ 
(1) & (2) & (3) & (4) & (5) & (6) & (7) & (8) & (9) & (10) \\
\hline
 26.00& 5393.167&  3.24&  $-$0.91&  7.76& 140.3& 157.4&  $-$3.43&  2.204& +0.057\\
 26.00& 5395.215&  4.45&  $-$2.17&  7.91&  16.6&  22.3&  $-$5.73&  1.387& +0.154\\
 26.00& 5398.277&  4.45&  $-$0.67&  7.57&  70.8&  76.9&  $-$2.29&  1.891& +0.038\\
 26.00& 5401.264&  4.32&  $-$1.92&  7.68&  21.5&  27.1&  $-$5.50&  1.477& +0.135\\
 26.00& 5406.770&  4.37&  $-$1.72&  7.77&  32.0&  38.6&  $-$4.52&  1.631& +0.113\\
 26.00& 5409.133&  4.37&  $-$1.30&  7.71&  49.7&  54.8&  $-$3.26&  1.771& +0.065\\
\hline
\end{tabular}
\end{center}
\end{table}

%Table 3 [6/10]
%\clearpage
\setcounter{table}{2}
\begin{table}[h]
\tiny
%\scriptsize
%\small
\caption{(Continued.) [6/10]}
\begin{center}
\begin{tabular}
{cccccrrrcc}\hline\hline
s-code& $\lambda$ & $\chi_{\rm low}$ & $\log gf$& $\log\epsilon_{00}$ &$W_{00}$ & $W_{31}$ & $K_{00}$ & $\alpha$ & $\beta$ \\ 
(1) & (2) & (3) & (4) & (5) & (6) & (7) & (8) & (9) & (10) \\
\hline
 26.00& 5696.102&  4.55&  $-$1.99&  7.50&   9.1&  11.5&  $-$6.00&  1.134& +0.166\\
 26.00& 5701.545&  2.56&  $-$2.22&  7.52&  80.9&  92.1&  $-$3.11&  1.983& +0.077\\
 26.00& 5705.466&  4.30&  $-$1.60&  7.63&  34.9&  39.8&  $-$4.64&  1.634& +0.092\\
 26.00& 5731.762&  4.26&  $-$1.30&  7.65&  53.6&  60.6&  $-$3.34&  1.805& +0.079\\
 26.00& 5732.275&  4.99&  $-$1.56&  7.61&  11.9&  15.1&  $-$5.34&  1.250& +0.177\\
 26.00& 5752.023&  4.55&  $-$1.27&  7.82&  49.7&  55.1&  $-$3.26&  1.758& +0.061\\
\hline
\end{tabular}
\end{center}
\end{table}

%Table 3 [7/10]
%\clearpage
\setcounter{table}{2}
\begin{table}[h]
\tiny
%\scriptsize
%\small
\caption{(Continued.) [7/10]}
\begin{center}
\begin{tabular}
{cccccrrrcc}\hline\hline
s-code& $\lambda$ & $\chi_{\rm low}$ & $\log gf$& $\log\epsilon_{00}$ &$W_{00}$ & $W_{31}$ & $K_{00}$ & $\alpha$ & $\beta$ \\ 
(1) & (2) & (3) & (4) & (5) & (6) & (7) & (8) & (9) & (10) \\
\hline
 26.00& 6157.725&  4.08&  $-$1.26&  7.51&  57.7&  65.6&  $-$3.21&  1.834& +0.049\\
 26.00& 6159.368&  4.61&  $-$1.97&  7.58&  10.6&  14.9&  $-$6.00&  1.236& +0.176\\
 26.00& 6165.361&  4.14&  $-$1.55&  7.54&  41.1&  49.1&  $-$4.29&  1.716& +0.072\\
 26.00& 6173.341&  2.22&  $-$2.88&  7.44&  64.7&  80.2&  $-$4.34&  1.920& +0.090\\
 26.00& 6180.203&  2.73&  $-$2.78&  7.50&  47.1&  59.8&  $-$5.47&  1.803& +0.107\\
 26.00& 6187.398&  2.83&  $-$4.34&  7.67&   3.0&   5.1&  $-$9.63&  0.804& +0.295\\
\hline
\end{tabular}
\end{center}
\end{table}

%Table 3 [8/10]
%\clearpage
\setcounter{table}{2}
\begin{table}[h]
\tiny
%\scriptsize
%\small
\caption{(Continued.) [8/10]}
\begin{center}
\begin{tabular}
{cccccrrrcc}\hline\hline
s-code& $\lambda$ & $\chi_{\rm low}$ & $\log gf$& $\log\epsilon_{00}$ &$W_{00}$ & $W_{31}$ & $K_{00}$ & $\alpha$ & $\beta$ \\ 
(1) & (2) & (3) & (4) & (5) & (6) & (7) & (8) & (9) & (10) \\
\hline
 26.00& 6710.316&  1.49&  $-$4.88&  7.49&  11.4&  20.1& $-$11.86&  1.359& +0.300\\
 26.00& 6713.771&  4.79&  $-$1.60&  7.66&  18.9&  21.9&  $-$5.06&  1.377& +0.095\\
 26.00& 6725.353&  4.10&  $-$2.30&  7.59&  15.2&  20.4&  $-$6.42&  1.359& +0.173\\
 26.00& 6726.661&  4.61&  $-$0.83&  7.27&  44.9&  47.7&  $-$3.73&  1.696& +0.041\\
 26.00& 6732.070&  4.58&  $-$2.21&  7.53&   6.3&   8.6&  $-$6.42&  0.982& +0.185\\
 26.00& 6733.151&  4.64&  $-$1.58&  7.62&  23.7&  28.7&  $-$4.99&  1.496& +0.119\\
\hline
\end{tabular}
\end{center}
\end{table}

%Table 3 [9/10]
%\clearpage
\setcounter{table}{2}
\begin{table}[h]
\tiny
%\scriptsize
%\small
\caption{(Continued.) [9/10]}
\begin{center}
\begin{tabular}
{cccccrrrcc}\hline\hline
s-code& $\lambda$ & $\chi_{\rm low}$ & $\log gf$& $\log\epsilon_{00}$ &$W_{00}$ & $W_{31}$ & $K_{00}$ & $\alpha$ & $\beta$ \\ 
(1) & (2) & (3) & (4) & (5) & (6) & (7) & (8) & (9) & (10) \\
\hline
 27.00& 5342.695&  4.02&  +0.69&  4.83&  28.3&  30.8&  $-$4.90&  1.542& +0.083\\
 27.00& 5359.192&  4.15&  +0.34&  4.67&   8.8&  12.0&  $-$6.20&  1.130& +0.177\\
 27.00& 5381.768&  4.24&  $-$0.03&  4.92&   5.7&   8.1&  $-$6.08&  0.934& +0.169\\
 27.00& 5454.572&  4.07&  $-$0.24&  5.28&  11.1&  14.1&  $-$6.25&  1.197& +0.154\\
 27.00& 6086.658&  3.41&  $-$1.04&  4.58&   1.8&   4.6&  $-$9.63&  0.719& +0.420\\
 27.00& 6093.143&  1.74&  $-$2.44&  4.97&   6.9&  15.0& $-$11.23&  1.245& +0.363\\
\hline
\end{tabular}
\end{center}
\end{table}

%Table 3 [10/10]
%\clearpage
\setcounter{table}{2}
\begin{table}[h]
\tiny
%\scriptsize
%\small
\caption{(Continued.) [10/10]}
\begin{center}
\begin{tabular}
{cccccrrrcc}\hline\hline
s-code& $\lambda$ & $\chi_{\rm low}$ & $\log gf$& $\log\epsilon_{00}$ &$W_{00}$ & $W_{31}$ & $K_{00}$ & $\alpha$ & $\beta$ \\ 
(1) & (2) & (3) & (4) & (5) & (6) & (7) & (8) & (9) & (10) \\
\hline
 28.00& 6861.236&  5.36&  $-$0.72&  6.18&   3.7&   4.0&  $-$3.96&  0.628& +0.031\\
\hline
\multicolumn{10}{c}{[Cu~{\sc i} line, $\chi^{\rm ion}$ = 7.73~eV, minor population stage]} \\
 29.00& 5218.197&  3.82&  +0.48&  3.99&  49.3&  53.6&  $-$3.23&  1.752& +0.062\\
\hline
\multicolumn{10}{c}{[Zn~{\sc i} line, $\chi^{\rm ion}$ = 9.39~eV, neither major nor minor population]} \\
 30.00& 6362.338&  5.80&  +0.15&  4.66&  25.8&  25.8&  +0.56&  1.452& +0.033\\
\hline
\multicolumn{10}{c}{[Y~{\sc ii} lines, $\chi^{\rm ion}$ =12.24~eV, major population stage]} \\
\hline
\end{tabular}
\end{center}
%\scriptsize
\tiny
The value of ionization potential (in eV) and classification of
minor/major population stage are presented in the beginning of the section of each species.  
(1) Species code used in Kurucz's (1993) ATLAS9/WIDTH9 program 
({\it e.g.}, 6.00 for C~{\sc i}, 26.01 for Fe~{\sc ii}). (2) Wavelength (in \AA). 
(3) Excitation potential of the lower level (in eV).
(4) Value of $\log gf$ (dex). (5) Logarithmic abundance solution (in dex) 
at the disk center (in the usual normalization of $\log\epsilon_{\rm H} = 12.00$).
(6) Equivalent width at the disk center (in m\AA).
(7) Equivalent width at the point nearest to the limb (in m\AA).
(8) Temperature-sensitivity index of at the disk center ({\it cf.} Equation~6).
(9) Constant of linear-regression relation for $\log W$ {\it vs.} $\mu$ ({\it cf.} Equation~7).
(10) Gradient of linear-regression line for $\log W$ {\it vs.} $\mu$ ({\it cf.} Equation~7). 
\end{table}

%Table 4
%\clearpage
\setcounter{table}{3}
\begin{table}[h]
\tiny
%\scriptsize
%\small
\caption{Byte-by-Byte Description of ``tableE.dat''.}
\begin{center}
\begin{tabular}
{ccccl}\hline\hline
 Bytes &  Format & Units &  Item &  Brief Explanations \\
\hline
~~1--~12 &  A12  &  ---  &   line-code &  spectral line code$^{(a)}$ \\
~13--~18 &  F6.2 &  eV   &   $\chi^{\rm ion}$   &  ionization potential from the ground level$^{(b)}$\\
~20--~20 &  A1   &  ---  &   class    &  whether the species is minor or major population stage$^{(c)}$\\
~22--~26 &  F5.2 &  ---  &   s-code    &  constructed from atomic number and ionization degree$^{(d)}$\\ 
~28--~35 &  F8.3 & \AA\  &   $\lambda$&  line wavelength$^{(e)}$ \\
~36--~41 &  F6.3 &  eV   &   $\chi_{\rm low}$   &  lower excitation potential$^{(e)}$\\
~43--~49 &  F7.3 &  dex  &   $\log gf$    & $g$ is stat. weight of lower level and $f$ is osc. strength.$^{(e)}$\\
~51--~56 &  F6.3 &  dex  &   $\log\epsilon_{00}$&  best-fit abundance$^{(f)}$ solution at point~00 (disk center)\\
~57--~63 &  F7.2 & m\AA\ &   $W_{00}$      &  equivalent width at point~00 (disk center)\\
~64--~70 &  F7.2 & m\AA\ &   $W_{31}$      &  equivalent width at point~31 (nearest to the limb)\\
~71--~77 &  F7.2 &  ---  &   $K_{00}$      &  $d\log W/d\log T$ ($T$-sensitivity parameter) for disk center\\
~79--~85 &  F7.2 &  dex  &   $\alpha$    &  constant of linear-regression relation ({\it cf.} Equation~7) \\
~86--~92 &  F7.2 &  dex  &   $\beta$     &  gradient of linear-regression relation ({\it cf.} Equation~7) \\
~93--~98 &  F6.2 &  dex  &   $gammar$   &  radiation damping parameter$^{(g)}$ \\
~99--104 &  F6.2 &  dex  &   $gammas$   &  Stark effect damping parameter$^{(g)}$\\
105--110 &  F6.2 &  dex  &   $gammaw$   &  van der Waals effect damping parameter$^{(g)}$\\
112--127 &  A16  &  ---  &   t-info(l)   &  term information of lower level$^{(h)}$\\
129--144 &  A16  &  ---  &   t-info(u)   &  term information of upper level$^{(h)}$\\
145--152 &  F8.3 &  ---  &   $g^{\rm L}_{\rm eff}$  &  effective Lande's $g$-factor$^{(i)}$\\
153--160 &  F8.2 & \AA\  &   $\lambda_{1}$  &  shorter wavelength limit of profile-fitting range\\ 
161--168 &  F8.2 & \AA\  &   $\lambda_{2}$  &  longer wavelength limit of profile-fitting range\\
171--175 &   A5  &  ---  &   w-header  &  spectrum header used for this line ({\it cf.} Table~2)\\
\hline
\end{tabular}
\end{center}
%\small
%\scriptsize
\tiny

Notes:\\
$^{(a)}$Constructed from the species code and wavelength. For example:
     O~{\sc i}  line (species code:  8.00) at 5577.339~\AA\ $\rightarrow$ 0800\_5577339 is the line-code.
     Fe~{\sc i} line (species code: 26.00) at 5123.719~\AA\ $\rightarrow$ 2600\_5123719 is the line-code.
     Y~{\sc ii} line (species code: 39.01) at 6795.414~\AA\ $\rightarrow$ 3901\_6795414 is the line-code.\\
$^{(b)}$Taken from the data of Kurucz's (1993) WIDTH9 program.\\
$^{(c)}$Three cases are possible: m...minor population stage, M...major population stage, x...indefinite.\\
$^{(d)}$For example: C~{\sc i} $\rightarrow$ 6.00, Fe~{\sc ii} $\rightarrow$ 26.01, Ni~{\sc i} $\rightarrow$ 28.00.\\
$^{(e)}$Taken from Kurucz and Bell (1995).\\
$^{(f)}$Abundances are expressed in the usual definition of $\log\epsilon = \log(N/N_{\rm H})+ 12$.\\
$^{(g)}$$gammar$: logarithm of radiation damping width (s$^{-1}$) [$\log\gamma_{\rm rad}$]. 
       $gammas$: logarithm of Stark damping width (s$^{-1}$) per electron density (cm$^{-3}$) 
       at 10000 K [$\log(\gamma_{\rm e}/N_{\rm e})$].
       $gammaw$: logarithm of van der Waals damping width ($s^{-1}$) per hydrogen density 
       (cm$^{-3}$) at 10000 K [$\log(\gamma_{\rm w}/N_{\rm H})$].
     These data were taken from the compilation of Kurucz and Bell (1995).
     In case that no data are available therein (indicated as the dummy values of 0.00 
     in this table), we used the default values computed by Kurucz's (1993) WIDTH9 program.\\
$^{(h)}$This atomic level information was taken from Kurucz and Bell (1995),
     which includes $J$ (quantum number of total angular momentum) as well as
     the spectral term of LS coupling ($2S+1$ and S, P, D, ... , from which the spin 
     quantum number $S$ and orbital angular momentum quantum number $L$ can be derived). \\
$^{(i)}$This effective Lande factor was derived from the $L$, $S$, and $J$ values of both 
     the upper and lower level in the conventional manner (see, {\it e.g.}, Gray 1988).
     In the indeterminable case where the necessary information is lacking, a dummy value
     of $-9.999$ is given.
\end{table}

%Table 5
%\clearpage
\setcounter{table}{4}
\begin{table}[h]
\tiny
%\scriptsize
%\small
\caption{Byte-by-Byte Description of ``????\_???????.dat''.}
\begin{center}
\begin{tabular}
{ccccl}\hline\hline
 Bytes &  Format & Units &  Item &  Brief Explanations \\
\hline
 ~~1--~~2 &   A2  &  --- &     $i$  & No. of observed point on the solar disk (00, 01, ...31) \\
 ~~3--~10 &  F8.4 &  --- &   $\sin\theta_{i}$ &  sine of angle $\theta^{(*)}$ at point $i$ \\
 ~11--~18 &  F8.4 &  --- &   $\cos\theta_{i}$ &  cosine of angle $\theta^{(*)}$ at point $i$\\
 ~19--~25 &  F7.3 &  dex &   $\log\epsilon_{i}$ & abundance solution at point $i$ \\
 ~26--~32 &  F7.3 &  dex &   $\log W_{i}$  &  logarithm of equivalent width (m\AA) at point $i$ \\
 ~33--~39 &  F7.2 &  m\AA\  &    $W_{i}$   &   equivalent width at point $i$ \\
 ~40--~46 &  F7.3 &  dex & $\langle \log\tau_{i}\rangle$ &  mean line-formation depth at point $i$ \\ 
 ~47--~53 &  F7.3 & km~s$^{-1}$ & $V_{{\rm los},i}$ & line-of-sight velocity dispersion at point $i$ \\
\hline
\end{tabular}
\end{center}
%\small
%\scriptsize
\tiny
The header of the filename ``????\_???????'' is the line-code ({\it cf.} the note in Table~4).\\
$^{(*)}$Angle between the line of sight and the normal to the surface.
\end{table}
%\end{article} 
\end{document}